\documentclass[pra,floatfix,showpacs,twocolumn,superscriptaddress]{revtex4}
\usepackage{amsmath}
\usepackage{epsfig,graphicx,times}

\newcommand{\be}{\begin{equation}}
\newcommand{\ee}{\end{equation}}
\newcommand{\bea}{\begin{eqnarray}}
\newcommand{\eea}{\end{eqnarray}}
\providecommand{\U}[1]{\protect\rule{.1in}{.1in}}
\begin{document}

\title{Fast ground-state cooling of mechanical resonator with time-dependent optical cavities}
\author{Yong Li}
\affiliation{Department of Physics and Center of Theoretical and Computational Physics,
The University of Hong Kong, Pokfulam Road, Hong Kong, China} \affiliation{Beijing
Computational Science Research Center, Beijing 100084, China}
\author{Lian-Ao Wu}
\affiliation{Department of Theoretical Physics and History of Science, The Basque Country
University (EHU/UPV), P.O. Box 644, 48080 Bilbao, Spain} \affiliation{IKERBASQUE, Basque
Foundation for Science, 48011 Bilbao, Spain}
\author{Z. D. Wang}
\affiliation{Department of Physics and Center of Theoretical and Computational Physics,
The University of Hong Kong, Pokfulam Road, Hong Kong, China}
\date{\today }
\pacs{42.50.Vk, 03.65.-w, 42.50.Dv}

\begin{abstract}
We propose a feasible scheme to cool down a mechanical resonator (MR) in a
three-mirror cavity optomechanical system with controllable external optical drives.
Under the Born-Oppenheimer (BO) approximation, the whole dynamics of the mechanical
resonator and cavities is reduced to that of a time-dependent harmonic oscillator,
whose effective frequency can be controlled through the optical driving fields. The
{fast} cooling of the MR can be realized by controlling the amplitude of the optical
drives. Significantly, {we further show that the ground-state cooling may be achieved
via the three-mirror cavity optomechanical system without the resolved sideband
condition.}
\end{abstract}

\maketitle

\pagenumbering{arabic}


\section{Introduction}

Ground-state cooling of nano-mechanical resonators (MRs) has attracted great
interests, as it is crucial in the improvement of detection precision of
MRs~\cite{Braginsky,ultrahigh detection}, observation of quantum behaviors of
macroscopic MRs~\cite{quantum-mechanical}, and quantum information processing based
on MRs~\cite{QIP-MR}. Over the years, methods to optimize the cooling of MRs in
optomechanical systems~\cite
{Metzger:2004,Gigan:2006,Arcizet:2006,Thompson:2007,Schliesser:2008,
Mancini:1998,Wilson-Rae:2007, Marquardt:2007, Li-Cooling,Kippenberg2007,
cooling-in-cavity,cooling-in-TLR} (or in electromechanical systems~\cite
{electromechanical}) have been studied extensively. In these cooling schemes, the MR
is coupled to a driven cavity through the optical radiation pressure and is cooled
via the passive backaction cooling (also called self-cooling)~ \cite
{Gigan:2006,Arcizet:2006,Wilson-Rae:2007,Marquardt:2007,Li-Cooling,Kippenberg2007}
without a feedback loop. More significantly, the backaction ground-state cooling of
MR~\cite{Wilson-Rae:2007,Marquardt:2007,Li-Cooling}, where the mean thermal
occupation number of phonons $\bar{n}$ is less than $1$, can be achieved when the
resolved sideband condition is satisfied.

Several experiments have reported progress in the backaction cooling of MR in
optomechanical systems~\cite{Metzger:2004,Gigan:2006,Arcizet:2006,
Thompson:2007,Schliesser:2008,cooling-in-cavity,cooling-in-TLR}, in particular those
approaching ground states of MRs. Recent research shows that MRs can be cooled to
states with a mean thermal occupation number $\bar{n}=35$ in an optical
cavity~\cite{cooling-in-cavity}, and $\bar{n}=3.8$ in a superconducting transmission
line resonator~\cite{cooling-in-TLR}.

Most existing models for ground-state cooling of MR in optomechanical systems use
resolved sideband cooling. The thermal phonon number of the MR is reduced by the
interaction between the MR and high-frequency auxiliary systems until the whole
system eventually reaches equilibrium. Here we propose an alternative but efficient
cooling method for MR in a three-mirror cavity (3MC) optomechanical
system~\cite{3MC}. We notice that the MR and its auxiliary system, e.g., optical
cavities, can be treated separately by the Born-Oppenheimer (BO)
approximation~\cite{BOA}. The MR therefore behaves as a single-mode harmonic
oscillator whose effective frequency is determined by the external optical driving
fields. By controlling the amplitude of optical driving fields, we can quickly reduce
the frequency of the bare MR, e.g., initially in a thermal equilibrium state, to a
smaller effective frequency, but retaining the populations. In other words, the MR is
cooled in a shorter time by doing work on the external system. Remarkably, by
combining such a fast cooling scheme with another process, we find that it is
feasible to achieve the \emph{ground-state cooling} of the MR with final effective
frequency that is the same as the bare one.

Our ground-state cooling mechanism is distinctly different from that of the
conventional sideband cooling in optomechanical systems: (i) Realization of sideband
cooling requires a long time until a steady state is reached. Here the MR {may} be
cooled down to a non-steady state at a short time and may later become hotter again
because of interactions with environment. (ii) The optical driving field is
time-independent in the sideband cooling, while the present scheme needs a
time-dependent and much stronger optical power. (iii) The sideband cooling of MR
happens when the optical detuning $\Delta $ approaches the bare frequency of MR
$\omega $, {the cooling condition here is far detuned from the cavity ($|\Delta |\gg
\omega $), which greatly simplifies the requirement for experimental control of the
optical detuning. (iv) The decay rate of cavities in our scheme can be larger than
the frequency of MR, while it is not allowed in resolved sideband cooling. }


\section{Three-mirror cavity configuration}

\begin{figure}[th]
\includegraphics[width=7.8cm]{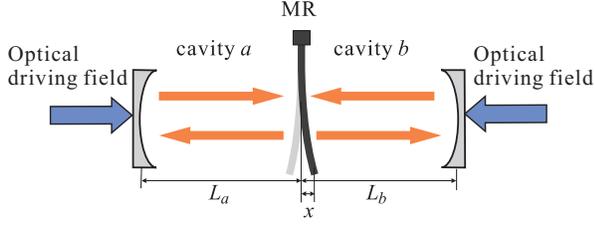}
\caption{(Color online) The schematics of 3MC optomechanical system. The movable MR
with two perfectly reflecting surfaces is placed inside a driven cavity with two
transmissive fixed mirrors.} \label{3mc}
\end{figure}

Unlike the conventional two-mirror cavity (2MC) optomechanical system, we consider
here an optomechanical system of 3MC configuration, as shown in Fig.~\ref{3mc}. The
mirror of MR with two perfectly reflecting surfaces is placed inside a cavity and the
two other fixed mirrors are transmissive and subject to external optical drives. This
setup is different from that in Ref.~\cite{Thompson:2007}, where the MR is
transmissive.

The Hamiltonian of 3MC system can be written as
\begin{equation}
H=H_{c}+H_{s},  \label{H-ori}
\end{equation}
where $H_{s}=p^{2}/2m+m\omega ^{2}x^{2}/2$ is the Hamiltonian for a MR with effective
mass $m$ and bare eigenfrequency $\omega $, and
\begin{eqnarray}
H_{c} &=&\hbar (\omega _{a}-G_{a}x)a^{\dag }a+\hbar (\omega
_{b}+G_{b}x)b^{\dag }b  \notag \\
&&+\hbar (\xi _{a}e^{i\nu _{a}t}a+\xi _{b}e^{i\nu _{b}t}b+\text{H.c.})  \label{Hf}
\end{eqnarray}
describes the optomechanical coupling between single-mode optical driven cavities a
and b (with the eigenfrequencies $\omega _{a}$ and $\omega _{b}$, respectively) and
the MR. Here $\xi _{a,b}$ are the amplitudes of external optical driving fields to
the cavities with the corresponding drive frequencies $\nu _{a,b}$, $G_{a}=\omega
_{a}/L_{a}$ and $G_{b}=\omega _{b}/L_{b}$ are the corresponding optomechanical
coupling strengths via the radiation pressure, and $L_{a,b}$ are the lengths of the
cavities $a$ and $b$.

Similar to the case for electrons and nuclei in a molecule, the
eigenfrequencies $\omega _{a}$ and $\omega _{b}$ of two cavities are much
higher than the frequency $\omega $ of the MR {such that} the BO approximation
can be employed to separate the degrees of freedom of MR from those of two
cavities. For simplicity, but without loss of generality, we set $\omega
_{a}=\omega _{b}$, $G_{a}=G_{b}=G$, $L_{a}=L_{b}=L$, $\xi _{a}=\xi _{a}^{\ast
}=\xi _{b}=\xi _{b}^{\ast }=\xi $, and $\nu _{a}=\nu _{b}=\nu $. The cavity
Hamiltonian can now be written as
\begin{equation}
H_{c}=\hbar (\Delta -Gx)a^{\dag }a+\hbar (\Delta +Gx)b^{\dag }b+\hbar \xi
(a+b+\text{H.c.}) \label{H-c}
\end{equation}%
in a rotating frame with respect to $H_{f,0}=\hbar \nu (a^{\dag }a+b^{\dag }b)$,
where $\Delta \equiv (\omega _{a}-\nu )$ is the optical detuning at the absence of
the motion of the MR.

We define operators $\widetilde{a}=a-\alpha $ and $\widetilde{b}=b-\beta $,
where $\alpha =-\xi /(\Delta -Gx)$ and $\beta =-\xi /(\Delta +Gx)$, such
that the Hamiltonian~(\ref{H-c}) can be expressed as
\begin{equation}
H_{c}=\hbar (\Delta -Gx)\widetilde{a}^{\dag }\widetilde{a}+\hbar (\Delta +Gx)%
\widetilde{b}^{\dag }\widetilde{b}+c_{0}(x),  \label{H-c2}
\end{equation}%
where
\begin{equation*}
c_{0}(x)=\frac{-2\hbar \Delta \xi ^{2}}{\Delta ^{2}-G^{2}x^{2}}\approx \frac{%
-2\hbar \xi ^{2}}{\Delta }(1+\frac{\omega ^{2}}{\Delta ^{2}L^{2}}x^{2})
\end{equation*}%
is $x$ dependent. Note that we have used the condition $Gx/\Delta \ll 1$ and ignored
the effects from the decay rates ($\kappa _{a}=\kappa _{b}$ $=\kappa $) of cavity
fields based on the fact that $\kappa $ is much less than the detuning $\Delta $.

When the cavity fields are fixed for the state $|\widetilde{0}\rangle $ ($%
\equiv |\widetilde{0}(x)\rangle $) with $\widetilde{a}|\widetilde{0}\rangle
=0$ and $\widetilde{b}|\widetilde{0}\rangle =0$, the effective Hamiltonian
of the MR is
\begin{eqnarray}
H_{\mathrm{eff}} &=&\frac{(p+A_{\widetilde{0}})^{2}}{2m}+\frac{1}{2}m\omega
^{2}x^{2}+c_{0}(x)  \notag \\
&\equiv &\frac{p^{2}}{2m}+\frac{1}{2}m\omega _{\mathrm{eff}}^{2}x^{2}+c_{r},
\label{H-r-eff}
\end{eqnarray}%
where the effective frequency is given by 
$\omega _{\mathrm{eff}}^{2}=\omega ^{2}-4\hbar \xi ^{2}\omega
_{a}^{2}/(m\Delta ^{3}L^{2})$ 
and $c_{r}=-2\hbar ^{2}\xi ^{2}/\Delta $ is a constant shift. We have assumed that
the induced gauge potential $A_{\widetilde{0}}=-i\hbar \langle \widetilde{0}|\nabla
|\widetilde{0}\rangle $ $=-i\hbar \langle \widetilde{0} (x)|\frac{\partial }{\partial
x}|\widetilde{0}(x)\rangle =0$ if $|\widetilde{ 0}(x)\rangle $ {is} a real function.
Then the MR behaves as a harmonic oscillator with the effective frequency determined
by $\xi$ (equivalently the amplitudes of the optical driving fields), which is
similar to the discussion according to the conventional optical
springs~\cite{optical-spring}.

The 3MC configuration distinguishes itself from the 2MC configuration because the
high symmetry of the three-mirror cavity allows to significantly improve the trap
stiffness and to partially remove the bistability as addressed in Ref.~\cite {3MC}.
The MR displacement in the effective Hamiltonian~(\ref{H-r-eff}) is zero, while the
corresponding MR displacement is finite in the 2MC configuration, making the dynamics
of the 2MC more complicated.


\section{General fast cooling of time-dependent MR}

Although the effective Hamiltonian~(\ref{H-r-eff}) is obtained by using the
time-independent BO approximation for the time-independent $\xi $, the similar result
also holds for a time-dependent parameter $\xi (t)=\xi _{0}f(t)$ when the factor
$f(t)$ ($\left\vert f(t)\right\vert <1$) varies very slowly in comparison with the
detuning $\Delta ,$ i.e., $\dot{f}(t)/f(t)\ll \Delta $. In this case, the
Schr\"{o}dinger equation $H\left\vert \psi \right\rangle =i\hbar \partial
_{t}\left\vert \psi \right\rangle $ can be simplified by a similar BO approximation
with a highly non-trivial consideration of time dependency, as discussed in
Ref.~\cite{Rabitz92}. The BO approximation implies $\langle \widetilde{0}|H_{s}|\psi
\rangle \approx H_{s}\langle \widetilde{0}|\psi \rangle $ such that $\langle
\widetilde{0}|H|\psi \rangle \approx (H_{s}+c_{0}(x))\langle \widetilde{0} |\psi
\rangle .$ Here $\langle \widetilde{0}|\psi \rangle $ is an instantaneous eigen
wave-function of the MR with the effective Hamiltonian $ H_{c}$, $c_{0}(x)$ is the
corresponding instantaneous eigenvalues and now is time dependent,
$c_{0}(x)=c_{0}(x,t)$.

Based on the adiabatic theorem with conditions $\dot{f}(t)/f(t)\ll \Delta $
and the wave function $|\widetilde{0}\rangle $ being real, we obtain $%
\partial _{t}\langle \widetilde{0}|\approx 0$. The effective Schr\"{o}dinger
equation for the MR is then given by
\begin{equation}
[H_{s}+c_{0}(x,t)]\langle \widetilde{0}|\psi \rangle =i\hbar \partial
_{t}\langle \widetilde{0}|\psi \rangle .
\end{equation}%
Therefore, the MR can be described as a time-dependent harmonic oscillator with the
effective Hamiltonian
\begin{equation}
H_{\mathrm{eff}}(t)=H_{s}+c_{0}(x,t)=\frac{p^{2}}{2m}+\frac{1}{2}m\omega _{%
\mathrm{eff}}^{2}(t)x^{2}.  \label{BO2}
\end{equation}%
Here the corresponding time-dependent eigen-frequency
\begin{equation}
\omega _{\mathrm{eff}}(t)=\omega \sqrt{1+\eta f^{2}(t)}
\end{equation}%
with $\eta =-4\hbar \xi _{0}^{2}\omega _{a}^{2}/(m\omega ^{2}\Delta ^{3}L^{2})$, is
controlled by the external optical field via the dimensionless function $f(t)$.

The dynamics of a time-dependent harmonic oscillator is analytically solvable using
Lewis-Reisenfeld invariants~\cite{invariant-motion}. Here we are only interested in
the specific trajectories $\omega _{\mathrm{eff}}(t)$ wherein the instantaneous
populations at the initial time are the same as those at the final time, but the
eigenfrequency and the corresponding {average} energy at the final time decreases.
{That means the harmonic oscillator has the same entropy at the initial and final
times and} is cooled by doing work on the external field, instead of having heat
flowing out of it. Such a cooling trajectory 
{does} not depend on the initial state. The simplest way to realize the cooling
trajectory is the adiabatic process where the populations remain the same all the
time. The disadvantage is that the adiabatic process needs a long time, during which
the relaxation of the harmonic oscillator may bring negative effects. Alternatively,
an optimal bang-bang process has been proposed to achieve a cooling trajectory in a
shorter time \cite{bang-bang}. The effective frequency is real and stepwise function
of time in this process, where the final instantaneous populations of the MR remain
the same as the initial ones in terms of controlling the step values and durations of
the effective frequency. Recently a fast optimal frictionless
process~\cite{Chen2010,Schaff2010} was proposed to achieve atom cooling by choosing a
certain trajectory of effective frequency, wherein the instantaneous effective
frequency is allowed to be \textquotedblleft imaginary".

Employing these methods, one is able to cool the MR in time-controllable
optomechanical systems, as done in {atom cooling}~\cite{bang-bang,
Chen2010,Schaff2010}. Let us estimate the feasibility of cooling a MR with reasonable
empirical parameters: {the bare eigenfrequency of MR $\omega /2\pi =134$ kHz, the
effective mass $m=50$ pg~\cite{Thompson:2007}, the eigenfrequency of the optical
cavity $\omega _{a}/2\pi \approx 7\times 10^{14}$ Hz, the optical detuning $\Delta
/2\pi \approx 10^{7}$ Hz}, and the length of the optical cavity $L\approx 2$ mm. The
optical driving fields are controllable and can have the time-dependent form $\xi
=\xi _{0}f(t)$ {with $ \xi _{0}/2\pi \approx 10^{9}$ Hz} and $|f(t)|<1$. With these
parameters, we obtain {$\eta \approx -9$}. The effective frequency $\omega
_{\mathrm{eff} }(t)$ can be much smaller than the bare frequency $\omega $ of the MR,
or even be {\textquotedblleft imaginary" by tuning }$f(t)$.

Specifically, consider a MR in the thermal state with the mean phonon occupation
number $\bar{n}(t_{\mathrm{i}})=1/[\exp (\hbar \omega /k_{B}T)-1]$ $\approx
k_{B}T/\hbar \omega $ ($\gg 1$) at the initial time $t_{\mathrm{i}}$, and $\omega
_{\mathrm{eff}}(t_{\mathrm{i}})=\omega $. One can design a trajectory between $\omega
_{\mathrm{eff}}(t_{\mathrm{i}})$ and the much lower effective frequency $\omega
_{\mathrm{eff}}(t_{\mathrm{f}})$ at the final time (where $\omega
_{\mathrm{eff}}(t_{\mathrm{f}})$$\equiv \omega /R$ with $R\gg 1$ and the
corresponding $f(t_{\mathrm{f}})=\sqrt{(1-R^{-2})/\eta
}$) {such that the final mean} occupation number $\bar{n}(t_{\mathrm{f}})=%
\bar{n}(t_{\mathrm{i}})$. The energy of the MR is decreased by a factor $R$
after doing work on the external field, and the final effective temperature
is reduced to $T_{\mathrm{eff}}(t_{\mathrm{f}})\equiv \hbar \bar{n}(t_{\mathrm{f}%
})\omega _{\mathrm{eff}}(t_{\mathrm{f}})/k_{B}=T/R\ll T$.

We have proposed a fast cooling mechanism for the MR as a time-dependent effective
harmonic oscillator in a 3MC optomechanical system, where the final effective
frequency of the MR is different from the initial one. However, {the approach will
not lead to the ground-state cooling since the final mean population number of the
effective harmonic oscillator is the same as the original one which is usually much
larger than $1$.}

\section{Ground-state cooling of MR}

{In the preceding section, we obtain a fast cooling of MR wherein the final effective
frequency of the MR is different from the initial one. People may be mostly
interested in the cooling of the MR with the final effective frequency of the MR
being (nearly) equal to the original bare one, such as the sideband cooling of MR. In
this section, we design a different cooling approach for the same 3MC optomechanical
system} to cool the MR close to its ground state in a short time, with the final
effective frequency of the MR being equal to the original bare one.

We consider that the MR is originally (at $t=t_{\mathrm{o}}$) in a thermal
equilibrium state $\rho (t_{\mathrm{o}})=e^{-H_{\mathrm{eff}}(t_{\mathrm{o}
})/k_{B}T}/\mathrm{Tr}(e^{-H_{\mathrm{\mathrm{eff}}}(t_{\mathrm{o}})/k_{B}T}) $ in a
bath at a refrigeration temperature $T=20$ mK~\cite{note2} in the absence of the
optical driving fields. The corresponding mean occupation number of thermal phonons
is $\bar{n}(t_{\mathrm{o}})=$ $1/[\exp (\hbar \omega /k_{B}T)-1]\gg 1$. The MR is
then subject to the optical radiation pressure of the driven cavities, described by
the effective Hamiltonian in Eq.~(\ref{BO2}). We use the same preceding typical
parameters, except with the optical detuning $\Delta /2\pi =-10^{7}$ Hz and with the
maximum value of optical driving fields $\xi ^{(0)}/2\pi \approx 10^{12}$ Hz, and
obtain $\eta \approx 9\times 10^{6}$. At time $t=t_{\mathrm{i}}$ the MR evolves to a
new thermal state $\rho (t_{\mathrm{i}})=e^{-H_{\mathrm{eff}
}(t_{\mathrm{i}})/k_{B}T}/{\mathrm{Tr}(e^{-H_{\mathrm{eff}}(t_{\mathrm{i}
})/k_{B}T})}$ with the mean thermal occupation number $\bar{n}(t_{\mathrm{i}
})=1/[\exp (\hbar \omega _{\mathrm{eff}}(t_{\mathrm{i}})/k_{B}T)-1]$. We take
$f(t_{\mathrm{i}})=1$ such that $\omega _{\mathrm{eff}}(t_{\mathrm{i} })\approx
3000\omega $ and $\bar{n}(t_{\mathrm{i}})\approx 0.66<1\ll $ $\bar{
n}(t_{\mathrm{o}})\approx 3200$. The intermediate trajectory between $t_{
\mathrm{o}}$ and $t_{\mathrm{i}}$ may be arbitrary, on the condition that the MR is
in the thermal state at time $t_{\mathrm{i}}$.

{We need to design a special trajectory of the effective frequency $\omega
_{\mathrm{eff}}(t)$ from $t=t_{\mathrm{i}}$ to the final time $t=t_{\mathrm{f
}}$ such that $\omega _{\mathrm{eff}}(t_{\mathrm{f}})$ equals to the original
bare frequency $\omega $ and the corresponding state $\rho (t_{\mathrm{f}
})=\rho (t_{\mathrm{i}})$. Therefore, the final mean thermal occupation number
$\bar{ n}(t_{\mathrm{f}})=\bar{n}(t_{\mathrm{i}})\approx 0.66$, which indicates
that the MR is cooled near its ground state.}

\begin{figure}[th]
\includegraphics[width=6.6cm]{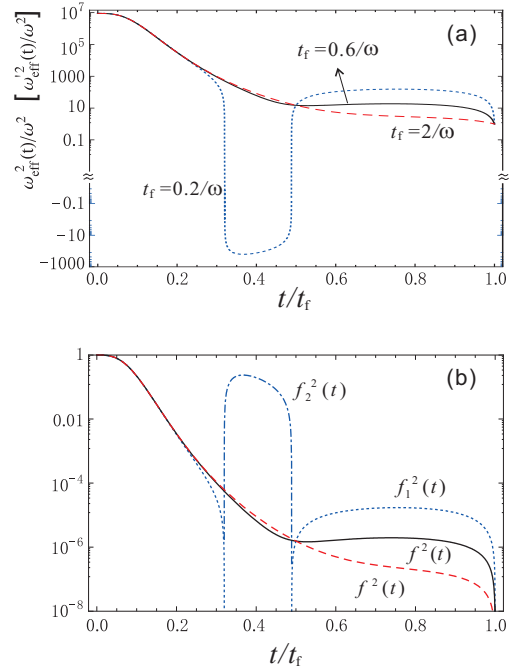}
\caption{(Color online) (a) The square of instantaneous eigen-frequency $
\protect\omega _{\mathrm{eff}}^{\prime }(t)$ (or $\protect\omega _{\mathrm{
eff}}(t)$) of MR from $\protect\omega _{\mathrm{\ eff}}^{\prime }(t_{\mathrm{
i}}=0)=3000\protect\omega $ to $\protect\omega _{\mathrm{eff}}^{\prime }(t_{
\mathrm{f}})=\protect\omega $ for different durations: $t_{\mathrm{f}}=0.2/
\protect\omega $ (blue dotted line); $t_{\mathrm{f}}=0.6/\protect\omega $ (black
solid line); $t_{\mathrm{f}}=2/\protect\omega $ (red dashed line). Here, the bare
frequency of MR is $\protect\omega =2\protect\pi \times 134$ kHz. The negative value
is due to an imaginary $ \omega _{\mathrm{eff}}^{\prime }(t)$ which appears for the
case of short evolution time $t_{\mathrm{f}}$. (b) The corresponding time-dependent
control parameters $f^{2}(t)$ [$f_{1}^{2}(t)$] (proportional to the optical powers)
and $f_{2}^{2}(t)$ (the blue dotted-dashed line, which contributes mainly to the
\textquotedblleft imaginary" $\protect\omega _{\mathrm{eff}}^{\prime }(t) $). }
\label{fig2}
\end{figure}

We now focus on designing such a trajectory of $\omega _{\mathrm{eff}}(t)$ in terms
of the Lewis-Riesenfeld invariant of motion~\cite{invariant-motion}. The invariant of
motion for our MR {harmonic oscillator} is $I(t)=m\omega
_{0}^{2}x^{2}/[2b^{2}(t)]+[b(t)p-m\dot{b}(t)x]^{2}/2$, where $\omega _{0}=\omega
_{\mathrm{eff}}(t_{\mathrm{i}})$. The dimensional real function $ b(t)$ satisfies the
condition $\ddot{b}(t)+\omega _{\mathrm{eff} }^{2}(t)b=\omega
_{0}^{2}/b^{3}(t)$~\cite{invariant-motion}, which is used in Ref.~\cite{Chen2010} for
cold atoms. We show that the simplest polynomial choice of $b(t)$~\cite{note} will
lead to an above-required cooling trajectory, which can be realized by controlling
the parameters $f(t)$, as shown by {the solid and dashed lines} in Fig.~\ref{fig2}.

In addition, the cooling process could even be faster if \emph{imaginary} frequencies
$\omega _{\mathrm{eff}}(t)$ could be allowed as done in Ref.~\cite{Chen2010}, where
the imaginary frequencies cooling scheme were first proposed. The single optical mode
Hamiltonian~(\ref{H-c2}) fails to provide such imaginary values of $\omega
_{\mathrm{eff}}(t)$ since $\eta $ is positive for the fixed parameters above. For
that purpose, we could adopt two optical modes in each cavity for the similar
symmetric 3MC configuration. A self-adjoint
\emph{Hamiltonian}~\cite{Note-self-adjoint} for the model with two optical modes in
each cavity in a rotating frame should be
\begin{eqnarray}
H_{c}^{\prime } &=&\hbar (\Delta _{1}-G_{1}x)a_{1}^{\dag }a_{1}+\hbar
(\Delta _{1}+G_{1}x)b_{1}^{\dag }b_{1}  \notag \\
&&+\hbar (\Delta _{2}-G_{2}x)a_{2}^{\dag }a_{2}+\hbar (\Delta
_{2}+G_{2}x)b_{2}^{\dag }b_{2}  \notag \\
&&+\hbar \lbrack \xi _{1}(a_{1}+b_{1})+\xi _{2}(a_{2}+b_{2})\text{H.c.}],
\end{eqnarray}%
where $a_{1,2}$ ($b_{1,2}$) denote the annihilation operators for the two modes in
cavity $a$ (cavity $b$) with the frequencies $\omega _{1,2}$. The coupling strengths
of optical radiation pressure are $G_{1,2}=\omega _{1,2}/L $. $\xi _{1}:=$ $\xi
_{1}^{(0)}f_{1}(t)$ and $\xi _{2}:=$ $\xi _{2}^{(0)}f_{2}(t)$ ($|f_{1,2}(t)|<1$) are
the amplitudes of the external time-dependent optical driving fields with frequencies
$\nu _{1,2}$. The optical detunings are $\Delta _{1}\equiv (\omega _{1}-\nu _{1})$
and $\Delta _{2}\equiv (\omega _{2}-\nu _{2})$.

Following the preceding discussion, the effective time-dependent Hamiltonian
for the MR becomes
\begin{equation}
H_{\mathrm{eff}}^{\prime }(t)\equiv \frac{p^{2}}{2m}+\frac{1}{2}m\omega _{%
\mathrm{eff}}^{\prime 2}(t)x^{2},  \label{H-r-eff-prime}
\end{equation}%
where the time-dependent eigenfrequency is given by
\begin{equation}
\omega _{\mathrm{eff}}^{\prime 2}(t)=\omega ^{2}[1+\eta
_{1}f_{1}^{2}(t)+\eta _{2}f_{2}^{2}(t)]  \label{frequency-eff-prime}
\end{equation}%
with two coefficients $\eta _{1}=-{4\hbar \xi _{1}^{(0)2}\omega _{1}^{2}}/{\
(m\omega ^{2}\Delta _{1}^{3}L^{2}})$ and $\eta _{2}=-{4\hbar \xi
_{2}^{(0)2}\omega _{2}^{2}}/{(m\omega ^{2}\Delta _{2}^{3}L^{2})}$.

We now have more effective control over the time-dependent eigenfrequency $\omega _{
\mathrm{eff}}^{\prime }(t)$ through two independent parameters $f_{1,2}(t)$
(proportional to the external input optical powers) and will use the same typical
parameters for both modes as before, {except for the optical frequencies $\omega
_{1}/2\pi \approx 7\times 10^{14}$ Hz, $ \omega _{2}/2\pi \approx 6\times 10^{14}$
Hz, the maximum amplitude of optical driving fields $\xi _{1}^{(0)}/2\pi \approx
10^{12}$ Hz, $\xi _{2}^{(0)}/2\pi \approx 10^{10}$ Hz, and the detunings $\Delta
_{1}/2\pi =-10^{7}$ Hz and $\Delta _{2}/2\pi =10^{7}$ Hz. Therefore, the coefficients
$\eta _{1}\approx 9\times 10^{6}$ and $\eta _{2}\approx -670$. Thus, the effective
frequency could be imaginary in the process when $f_{2}(t)$ is much larger than
$f_{1}(t)$ and the cooling can be achieved even faster (see the blue dotted and
dashed-dotted lines in Fig.~\ref{fig2})}.

{We wish to mention that our ground-state cooling scheme requires a strong input
optical power to obtain a strong modulation for the effective frequency. The maximum
value of the optical input power is $P=\hbar \omega _{a}{\xi^{(0)2} }/2\kappa \approx
1$ W for the maximum $\xi^{(0)} /2\pi \approx 10^{12}$ Hz, where the decay rate of
the optical modes of the cavities is assumed as $\kappa /2\pi \approx 10^{6}$ Hz. The
corresponding high finesse cavities with $F=\pi c/(2L\kappa )\approx 3.8\times
10^{4}$, are required.} We remark that the adiabatic condition for the time-dependent
BO approximation , $\dot{f}(t)/f(t)\ll |\Delta |$ and $\dot{f}_{1,2}(t)/f_{1,2}(t)\ll
|\Delta _{1,2}|$, is satisfied for the trajectories in Fig.~\ref{fig2}.

Our general analysis shows that we can realize the ground-state cooling for a MR with
the present approach. Even if the coupling between the MR and the optical cavity is
removed by shutting off the optical driving fields, the MR will remain in the same
ground state for a longer time after $t_{\mathrm{f}}$, until it approaches to a new
thermal equilibrium with its bath. During this period of time, one can perform
quantum operations, such as quantum information processing, on the MR. Note that the
controlled trajectory between $\omega _{\mathrm{eff}}(t_{\mathrm{i}})$ and $\omega _{
\mathrm{eff}}(t_{\mathrm{f}})$ is preformed in a short time of a fraction of
$1/\omega $, such that the relaxation process of the MR with a high Q factor (e.g.,
$10^{5}$) can be ignored.


\section{Conclusions}

We propose a feasible scheme to cool a MR through a 3MC optomechanical system. In the
limit of large optical detuning, we show that the optical fields can be eliminated
and the degree of freedom of the MR in the general Hamiltonian can be described by a
time-dependent harmonic oscillator. By controlling the amplitude of the external
input optical driving fields, we can manipulate the time-dependent effective
frequency and obtain the cooling of the MR via keeping the same populations of
instantaneous levels at the initial and final times during a special trajectory of
short time period. The ground-state cooling of MR can be realized as well. It is
encouraging that a similar trajectory has been implemented
experimentally~\cite{Schaff2010} for atom cooling. We believe that the same technique
is applicable to our scheme for a mechanical resonator.

\begin{acknowledgments}
We would like to thank Dr. P. Zhang for helpful discussions. This work was
supported by the RGC grant of Hong Kong No.~HKU7044/08P, the Ikerbasque
Foundation Start-up, the Spanish MEC No. FIS2009-12773-C02-02, and the State
Key Program for Basic Research of China (No.~2006CB921800).
\end{acknowledgments}

\end{document}